# Distributing Intelligence to the Edge and Beyond

Edgar Ramos, Roberto Morabito, and Jani-Pekka Kainulainen, Ericsson Research, Finland

**Abstract**

Machine Intelligence (MI) technologies have revolutionized the design and applications of computational intelligence systems, by introducing remarkable scientific and technological enhancements across domains. MI can improve Internet of Things (IoT) in several ways, such as optimizing the management of large volumes of data or improving automation and transmission in large-scale IoT deployments. When considering MI in the IoT context, MI services deployment must account for the latency demands and network bandwidth requirements. To this extent, moving the intelligence towards the IoT end-device aims to address such requirements and introduces the notion of Distributed MI (D-MI) also in the IoT context. However, current D-MI deployments are limited by the lack of MI interoperability. Currently, the intelligence is tightly bound to the application that exploits it, limiting the provisioning of that specific intelligence service to additional applications. The objective of this article is to propose a novel approach to cope with such constraints. It focuses on decoupling the intelligence from the application by revising the traditional device's stack and introducing an intelligence layer that provides services to the overlying application layer. This paradigm aims to provide final users with more control and accessibility of intelligence services by boosting providers' incentives to develop solutions that could theoretically reach any device. Based on the definition of this emerging paradigm, we explore several aspects related to the intelligence distribution and its impact in the whole MI ecosystem.

## I. Introduction

Artificial Intelligence (AI) has been one of the main objectives from the beginning of computer science. In fact, it is arguable that the ultimate objective of computer science is to replicate and perhaps outperform human intelligence.

Even though a significant amount of work has been done in AI through the years, it has been challenging to apply processing and memory-hungry techniques to our everyday applications until now. A great deal of AI has been deployed in data centers with large computing capabilities. In addition, specialized hardware has been created to manage certain AI models' specific processing needs, although the hardware has not fully reached mass production levels yet [1]. Nevertheless, the rise of the Internet of Things (IoT) has boosted the deployment of AI techniques to process the immense amount of data produced by always-connected devices. The resulting benefits of applying AI to IoT are not always totally understood; however, it is expected that bare computational power should be able to find valuable patterns or correlations that add to IoT devices' connectivity and monitoring services.

The actual characterization of AI is very broad. It can be defined differently whether the main concern is thought processes and reasoning behavior, or how closely AI should resemble humans or a pre-defined intelligence ideal. According to this concept, Russel and Norvig [2] organize the definitions of AI into four categories: thinking humanly, thinking rationally, acting humanly, and acting rationally. From a practical perspective and our everyday needs, humans might not need the full set of features and capabilities that AI can offer. Instead, humans require a sub-set of such capabilities that resembles intelligence at some level to support our everyday-life applications. This subset of features, which heavily relies on machine learning techniques and varies from case to case is what is known as Machine Intelligence (MI).

This article presents a vision for the distribution of MI. It is supported by the benefits of interoperability with the objective of democratizing MI's access. The paper is divided into five sections. The second section focuses on the relationship between MI and IoT, explores the current dominating models of intelligence provisioning, lists their limitations in the IoT context and looks at emerging trends. The third section presents a foresighted MI sharing ecosystem; it highlights the ecosystem's elements, roles, components, and functions as well as their requirements, possible implementations, and expectations. In particular, this section describes an intelligence layer that acts as a framework to connect all of the ecosystem components. The fourth section discusses the components' interactions related to intelligence provisioning and operation

Corresponding Author: Edgar Ramos (edgar.ramos@ericsson.com)

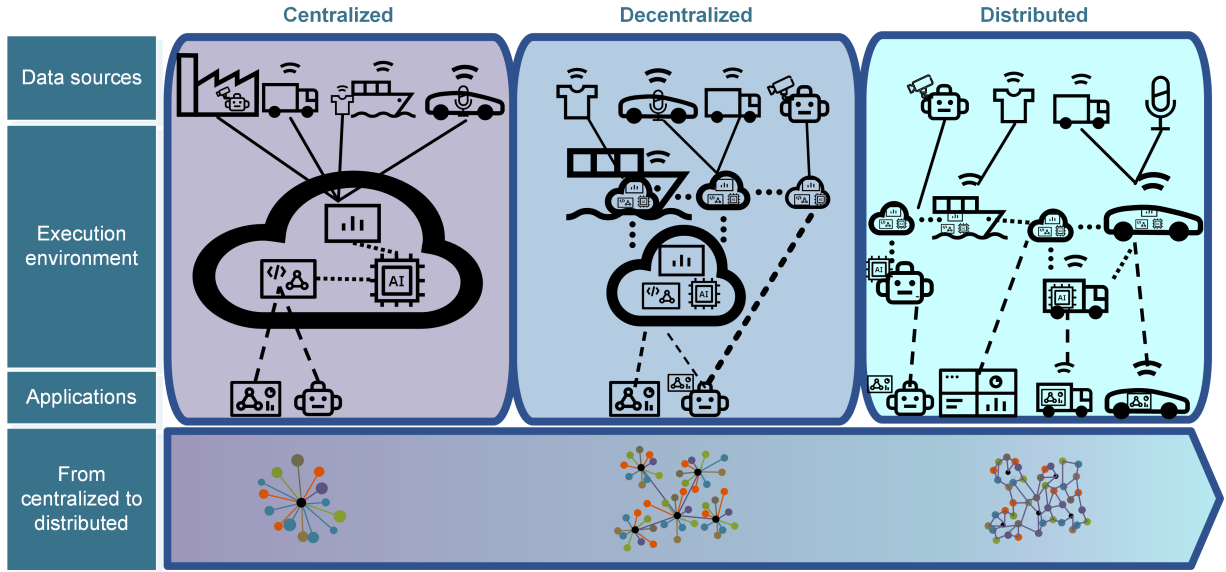

Fig. 1: From Centralized to Distributed machine intelligence deployments

environments. Furthermore, it also introduces additional technical enablers that support an interoperable Distributed MI (D-MI). The last section concludes the article by summarizing our vision of MI distribution and future work.

## II. Deployment of Machine Intelligence Systems and IoT

The IoT embraces the vision of an all connected world where any digital device can produce information that it is consumable locally and across locations. Experts estimate that by 2020, there will be between 15 to 30 billion connected IoT devices [3] [4] producing anywhere from 44 to 600 zettabytes (ZB) of data per year [5] [6]. Even if not all of this data is meant to be stored or captured for analysis, there will still be an immense quantity of data that requires automatic processing. Therefore, it is mainly machines, and not humans, who will interact with other machines to provide services, generate added value, and operate based on results from data-analytics. Processing machine-generated data includes functions such as optimization, prediction, anomaly detection, and error correction, which clearly fall under the MI umbrella. Consequently, MI systems will be one of the largest consumers, if not the largest consumer of IoT data in the future.

### A. Current Dominating Deployments

There are currently a few dominating models present in the intelligence space. Presently, the most common intelligence targeting IoT systems implementations are deployed in centralized solutions. Centralization is mainly due to the cost of training and adjusting the intelligent systems, as well as the required computational power. The applications and intelligence within them are mainly accessible through Application Programming Interfaces (APIs) via a network interface. Centralization also permits storing data that can be processed later by several intelligent algorithms. As a consequence, it enables the possibility to infer conclusions and draw correlations that serve as input to other intelligent systems in a seamless fashion. This approach seems to suit IoT data processing quite well since the collected data purpose is not immediately obvious at the moment.

As depicted in Figure 1, centralization manifests from several perspectives. From an execution perspective, the algorithms and models used are often executed (inference and/or training phases) in remote or local data centers. While it is not uncommon to see some intelligence in locally executed applications, many are frequently limited to very specific functions and are mostly based in proprietary models as part of the application itself. Another perspective is the MI Life Cycle Management (LCM). Generally, the provisioning, updating, substitution and decommission of MI is based on proprietary solutions. In many cases, these processes are performed based on agreements or as part of the application's own LCM. The MI used is like a black box; it is most fully understood by the creator entity. Some open source projects provide MI models that can be used freely (ex. [7], [8]), but these models frequently require a wrapping application and data science knowledge to be deployed. A final perspective is related to data. Most MI models require some period of training, adjusting, and testing until they can be used by an application. The entities that provision these models often provide the data required for training and adjusting them. These labeled data sets have become largely appreciated assets and, in many domains, a scarce resource. This shortage frequently causes the same data set to be reused across MI implementations, which means that the data's biases and possible defects are duplicated [9].

*B. The Typical Limitations of Dominant Deployments*

One valid question is why distribute intelligence processing. It is very intuitive to think that distribution is beneficial in certain environments such as local-data centric solutions, scenarios where the input of high volumes of data requires the use of online learning, and in connectivity constrained environments. However, it can be argued that centralized solutions where the data is collected and processed at a single point could be applied to almost any situation.

Data centers and cloud infrastructures seem to be able to grow their capabilities almost unlimitedly. And yet, in a world with a vision where everything that benefits from a connection will be connected (including shoelaces), one can ask if centralized processing of such a large volume of data is possible. Even if it was possible, would processing this data even be desirable or permissible by regulatory frameworks? And still, can centralization still remains feasible when requirements in terms of scalability and massive availability increase? As human beings, we tend to prefer having a sense of ownership of what affects our lives to give us the feeling of being in control. The ability to control, customize and limit our devices' intelligence is something very appealing to each of us. Therefore, distributing the intelligence to specific familiar points is likely more attractive than locating intelligence in an abstract virtual entity that exists in an unknown country data center. The last option is not a very pleasing thought to many people and makes them uneasy on the whole idea of MI. A relevant part of the privacy concerns is in part due to not well-established data governance models on collected and inferred data ownership and rights. In addition, it is not clear how to enforce the right to be forgotten in practice.

Another problem is the lack of interoperability of MI handling. Most often intelligence that is available for one application is not necessarily available for another; this specifically means that developers of the second application must build their own intelligence from scratch. This situation results in two scenarios. The first scenario involves only a few companies specializing in providing services for one type of application (ex. location-based suggestions or recommendations). This scenario is problematic because a small number of firms become the market leaders, dominating and dictating the existing offerings without considering interoperability. In the second scenario, a large number of scattered, diverse, and fragmented applications basically provide the same services. They often do not prioritize providing additional value to the intelligence collected, but mainly focus their efforts in trying to build the intelligence's functionality (ex. natural voice recognition). Therefore, the high fragmentation of solutions is the main interoperability challenge of this scenario. In conclusion, additionally to physical location, intelligence distribution must also refer to MI's ability to disseminate and access intelligence across multiple applications on the same device.

Autonomous transportation exemplifies how centralization can cause occasional problems. Centralization relies on network communication but, in some cases connectivity is not available 100% of the time. For example, autonomous ships would not be able to communicate with a central cloud for all the systems onboard. Another simple example depicting centralization problems is an application offering people detection from a video device. In the dominant type of deployment implementation, the training of the model is centrally done by the application developer or the device manufacturer, which often are the same entity. The manufacturer is also responsible for integrating MI into the application, maintaining the system, and controlling what type of intelligence is used within the device, in consequence limiting the possibility to add or modify intelligence services. For example, adding a service that recognizes not only humans, but in addition recognizes specific groups of people (like family members from a household) and provide special functions for them, it would not be possible in most of the cases. In addition, what happens if a device manufacturer goes out of business and the conditions of operation change? Similarly, what happens if the video content is of sensitive nature and the processing should be done locally instead of in a cloud? All these concerns cannot be easily addressed with centralization's current limitations.

*C. Shifting Towards a New Paradigm*

To make a more sustainable model, intelligence based on IoT data will likely require some degree of decentralization from a processing, storing, privacy, and probably a regulatory perspective. Determining the necessary extent of decentralization is crucial. As shown in Figure 1, on one end of the spectrum, there is a fully distributed system with topology and peer-to-peer transaction patterns; on the other end, there is a star-based, mainly client-server system. However, some intermediate options lie between these two alternatives. In a fully distributed system, part of centralization's simplicity and efficiency may become diluted in order to address network effects as well as reduce the risk of malicious attacks and local resources particularities. The decentralization of AI is being discussed and started to be explored in multiple contexts [10] [11] [12], and research about intelligence in the edge has already taken off [13] [14] [15].

Initiatives that pursue fully distributed systems are already being launched (SingularityNet[1], Ocean[2], OpenMined[3], etc) . These initiatives primarily focus on distributing AI model's processing and provisioning. They still maintain the centralized relationship between applications and the AI models, albeit it could be reduced by using Representational State Transfer (REST) APIs to access the model's services.

Currently, MI's LCM is not really being integrally addressed. For example, some works on lifelong learning [16] [17] focus on enabling adaptable systems to keep continuously learning and transfer learning from one instance to another, but do not focus in intelligence update. Service composition based on AI models' aggregation is also starting to emerge [18], including version compatibility management and that goes further in the right direction but still application's integration is left to implementation.

A model where intelligence is part of an end-to-end solution makes customization and intelligence sharing between applications demanding. It is also challenging to update such intelligence without changing the end-to-end solution. This difficulty renders intelligence services quite static and implementation-dependent. It tightly couples intelligence development and provision with the application developing roadmaps, the application providers limitations and interests, and their business strategies.

All the mentioned intelligence management challenges highlight that democratization of the MI is desirable. The systems can grant more power and flexibility to the end user by providing additional possibilities to decide what level of intelligence to use, what is the origin of the intelligence (both from data and processing perspective), which is the execution environment (local, distributed or network centralized) and what is done with the intelligence produced (transparency and privacy focus).

## III. The Distributed Machine Intelligence Ecosystem

*A. Interoperability in the Machine Intelligence Domain*

One of the key and more challenging aspects to promote interactions between ecosystem members is interoperability. This is particularly true for IoT ecosystems [19]. In the context of MI, the interoperability has to be considered at least in the following aspects:

**(i) MI LCM interoperability.** The LCMs of applications and MI are currently, if not the same, highly coupled. This level of coupling implies that changes in the intelligence have most of the time impact on the applications too. In order to make MI interoperable, it is imperative that their life cycle management is harmonized and possible to be managed by different entities. Figure 2 provides an example of MI's life cycle management indicating phases and their flow cycles. This interoperability aspect requires that any of the cycle operations could, in theory, be handled by exchangeable entities. If applied to our simple video processing example, one company could provide the model for performing the human recognition from the video feed. Another company could provide the data set for the recognition and train the model. A different company could add additional models or update the human recognition model to also provide family members recognition, and finally the service may be connected to a third party application that makes use of this capabilities to trigger locking and unlocking of doors next to the camera. Perhaps, in the future, the household camera might be used to recognize the family pet instead. The repurposing of the function implies that the human recognition model is not needed anymore and a new model to recognize the pet should be installed, restarting the cycle after decommissioning the initial model.

---

[1] https://singularitynet.io/   [2] https://oceanprotocol.com/   [3] https://www.openmined.org/

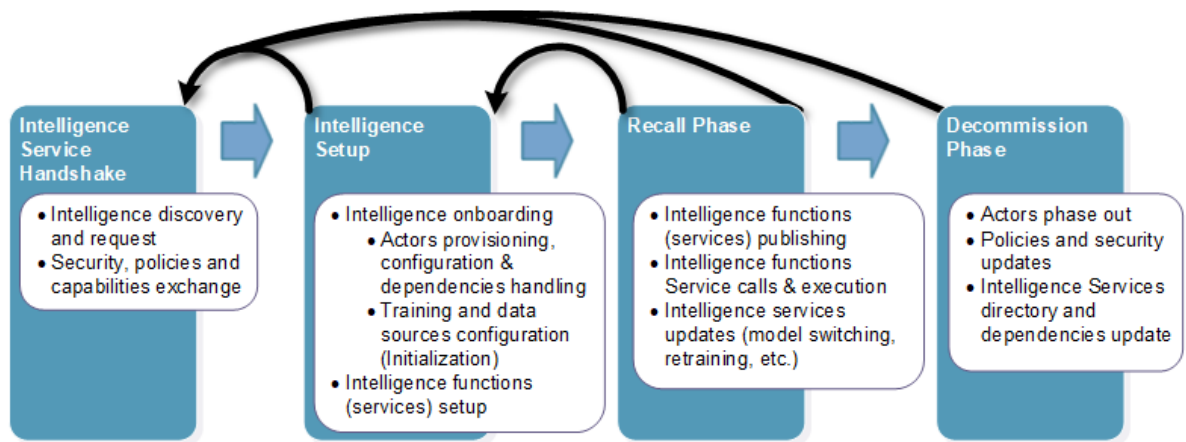

Fig. 2: Machine Intelligence Life Cycle Management

(ii) **Intelligent services interoperability.** The intelligent services are often a composition of several intelligent functions. Reusing one already accessible intelligent service and chaining it with others to generate a new service is key to promote value generation from the same resources that are already available for the systems. From the previous example, the model that recognizes the family members may be implemented as a service that takes as input the frames where the first model has recognized humans from the video. This new service is then attached to a third party application that matches the persons recognized to an application profile. To achieve this degree of integration, it is required to interoperate in the levels of service composition, defining how to chain the services together, understanding their functionality, and the nature of their required input and produced outputs. The same semantics required for the services composition can be used to associate the services to the applications thought connection points (such as APIs).

(iii) **Ecosystem's value sharing interoperability.** There are always ecosystem frictions that restrict the interactions between the actors or create unbalances that may end up in the ecosystem disintegration. Each of the ecosystem members has a value proposition to offer to the ecosystem, which makes resources available for other parties that at the same time help to generate additional value propositions. To promote and facilitate the integration it is needed the value sharing interoperability, where the benefits of the ecosystem can be exploited by all the parties involved. One example of a mechanism for value sharing interoperability is the Android Market. It provides a platform for developers to offer their value proposition in the form of an Android application, harmonizing at the same time the way how payments and trust between the rest of the players (device manufacturers, device users, service providers, etc) is handled. The other players also receive value from the market in the form of discovery of applications, reputation-based reviews, distribution channels and many other features depending on the role assumed in the whole Android ecosystem. The Android Market reduces the friction that otherwise could have existed between all the ecosystem players involved by providing ad-hoc interoperation of their value sharing.

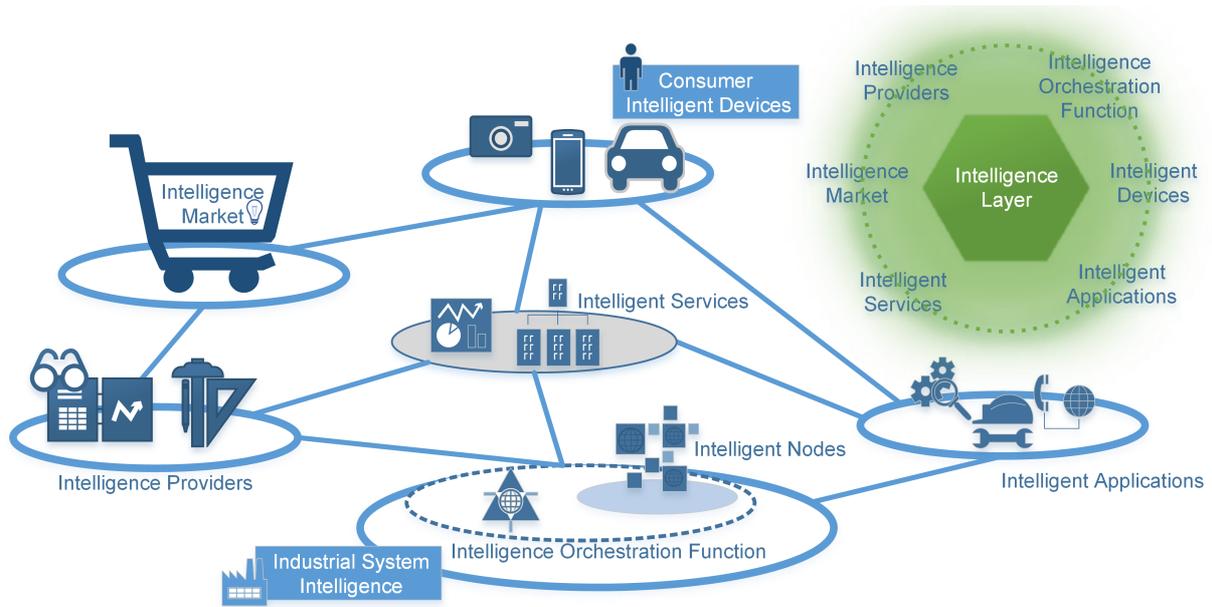

Fig. 3: Machine Intelligence Distribution Ecosystem

*B. Ecosystem Elements*

This section explores the MI sharing ecosystem's elements and places intelligence in a larger context. It is sometimes helpful to consider MI without thinking about a specific function or role that intelligence plays in various products. Instead, it is occasionally useful to think on intelligence's final goal and to whom it is directed. As a consequence, the need for an intelligence supply ecosystem with clearly defined roles and functions is realized. This need becomes even more evident when considering a D-MI environment where the intelligence is spread across several parts of the whole end-to-end application deployment.

In recent years, more powerful and portable hardware has emerged and intelligence has been brought closer to consumers. The applications could be fully executed in such hardware, or part of the heavy processing is offloaded to powerful centralized network environments. In order to efficiently utilize these capabilities and based on the previously presented considerations, Figure 3 depicts one possible ecosystem emerging to support MI distribution and the connections between its elements. The differences between the consumer and industrial sectors displayed will be further detailed in the following sections.

In this emergent ecosystem, the intelligence layer plays a key role. The intelligence layer whose definition and functionality will be extensively introduced in Section III-C serves as a framework that binds all the ecosystem's components. Basically, the layer abstracts the intelligence to a stratum domain where it can be addressed and manipulated.

In Figure 4, we provide a description of each of the ecosystem's component characteristics. Later in the article, some of these concepts will be further explained and contextualized. Moreover, we will discuss the considerations of including additional ecosystem components due to the particular differences between industrial and a consumer systems from a MI perspective.

*C. Intelligence Layer*

From a software engineering perspective, most applications that currently use MI integrate intelligence as part of the application itself. Alternatively, they may also interface with other applications in order to acquire the intelligence services required. This fact implies that MI is part of the application layer. It works well in tightly integrated systems and applications with well-defined boundaries that are not expected to change or be exposed to high variance. In such systems, the actual need for intelligence remain constant and the task and goal remains consistent over time. One example of this type of application is a chess-playing program. The rules and context of the game do not change and the goal is always the same.

| (Consumer) Intelligent Device | Intelligent Service |
|---|---|
| Device able to provide a wide set of intelligent functionalities and that can also take care of the processed intelligence actuation, by acting as a control loop and making possible to have very short reaction times. Through the definition of the intelligence layer, the device become a platform where the intelligence can be flexibly loaded, allowing a shift from a platform with defined purpose architecture to a machine intelligence optimized platform. | Service's characterization is tailored according to the level of granularity of the service. Specifically, we outline two different levels of service's granularity. An **Atomic Intelligence Service (AIS)** is a service that performs a very specific operation, the nature of which is not necessarily bound to the execution of an intelligent task; furthermore, it must not have any external dependency towards other AISs (e.g., normalization). A **Fine-Grained Intelligence Service (FGIS)** is composed by multiple AISs. From the observer's point of view, the composed FGIS provides intelligent functionality. This implies that at least one AIS component embeds an intelligent task (e.g., linear regression). |
| **Intelligence Market** | **Intelligent Application** |
| Enables the access to the intelligence from centralized rendezvous locations. It is a multi-sided platform that provide services of discovery, handshaking, subscription, delivery of intelligence and safe payments to users, as well as publishing, categorization, capabilities verification, secure connection and charging. The market may categorize intelligence according to target application, areas of scope, target platform, and any other possibility of grouping could be used to facilitate to users the discovery of the intelligence. | Located in the application layer, stands as the element that benefits of the output generated by the intelligent layer, by receiving as input the outcome of the intelligent service processing. |
| **Intelligence Provider** | **Intelligence Orchestration Function** |
| Defines, configures, and initializes the intelligence services from the intelligence layer. Furthermore, it handles intelligence algorithms and provides any data for initializing the local models (pre-computed models, training data, weights, and bias, etc.). Finally, it directs the user to other providers with specialized data or centralized processing capabilities according to the user requirements. | Executes a set of processes aimed at ensuring that the intelligence is suitably distributed and executed in the underlying system, taking due account the system architecture requirements. It also ensures an effective remote execution, update, and configuration of intelligent services, by making sure that the applications can properly receive and handle such services. In practical terms, it consists of at least two main components (**Intelligence Service manager** and **Intelligence Controller**), whose functionality are elucidated in Fig. 6 |

Fig. 4: Ecosystem Elements.

For environments where intelligence needs are constantly changing with less-defined functions or low coupling to the application architecture, the limitations of an integrated intelligence model are evident; to update the intelligence, the whole application needs to be updated. The most direct way to address this problem is to decouple intelligence from the application layer and make the intelligence as independent of it as possible. Then, the intelligence becomes a separated stratum that provides services to applications in the same way that other layers and platforms provide services (ex. the Operating System – OS).

An example can be visualized through an application that can detect a nearby dangerous object for a child. Traditionally, the application access the device's camera and GPS to take pictures, and applies object recognition algorithms matching the pictures with the location to detect nearby dangerous objects. If the application would be built on top of an intelligence layer, the application would ask the intelligence layer to provide notifications when dangerous objects are detected in the vicinity and, instead of detecting such objects by itself, it lets the intelligence layer to take care of such task. The application still must consider how to display the information and the level of detail required for the means of the application, but it does not need to consider object recognition processing part. Figure 5 compares the two described approaches related to how intelligence is integrated with the application layer.

This change in the architecture redefines the devices into platforms where the intelligence is loaded. Figure 6 attempts to outline and describe the additional key components that characterize the intelligence layer — Atomic Intelligence Service (AIS) and Fine-grained Intelligence Service (FGIS) are already defined in Figure 4—, based on MI provisioning requirements that will be further discussed in the following sections. In this context, an intelligent device is a device that supports the intelligence layered architecture and either fully or partially embeds the intelligence layer's components: intelligent services (with its sub-components AIS and FGIS), Intelligence Controller and Intelligence Service Manager. It is important to highlight that full or partial support depends on the device's computational capabilities. These capabilities can range from very constrained devices to high performance computing solutions, services availability, and the specific role conferred to the device within the entire ecosystem (e.g., edge/fog node, cloud instance, etc.). Figure 7 depicts a recurring scenario of many contemporary IoT deployments. The device space may consist of very constrained devices (ex. micro-controllers) and more capable devices (ex. single-board computers).

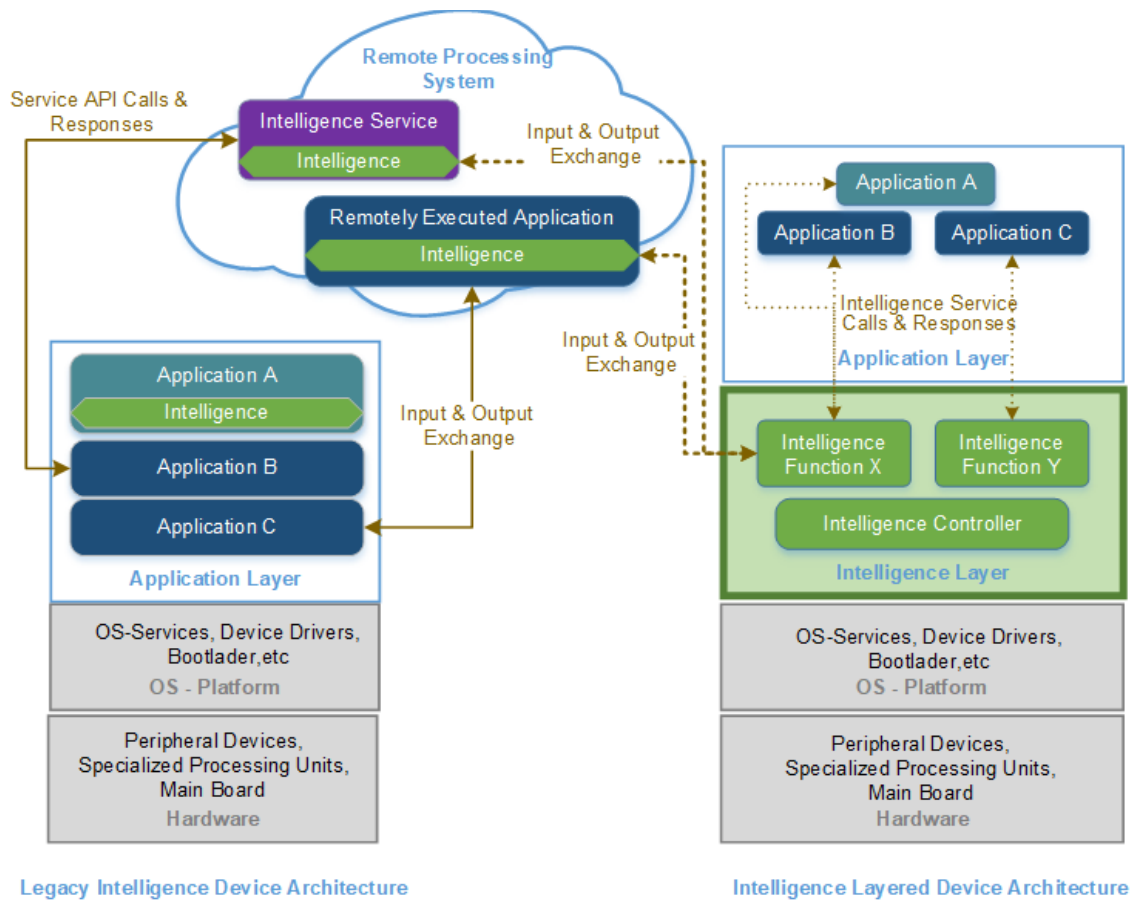

Fig. 5: Comparison of an intelligence-layered device architecture with a legacy intelligence device architecture.

As already previously introduced, the device's resource capabilities impact which intelligence layer components are implemented on top of it. For instance, for very constrained devices, it is reasonable to assume that intelligent services are hard-coded in the hardware; specifically, we refer to devices that make atomic intelligent services available out of the box. More capable devices can run more sophisticated services (ex. FGIS) and/or controlling mechanisms through the Intelligence Controller. The Intelligence Controller and the Intelligence Service Manager are required to execute many tasks. For example, a single device's Intelligence Controller must monitor local FGISs' execution and simultaneously exchange signaling information with other peers' controllers in order to effectively orchestrate and distribute intelligence. The Intelligence Controller must also be able to run local control loops and dynamically update the data flows (inbound and outbound). The Intelligence Service manager executes operations such as FGIS composition, security access enforcement and multi-purpose policies negotiation. Additionally, it accomplishes the crucial task of interfacing the intelligence and application layer by setting up and handling this purpose's specific functions. The FGIS may be hard-coded with a hardware implementation or controlled by the Intelligence Controller. The FGIS composition is performed by the Intelligence Service Manager's service composition functionality.

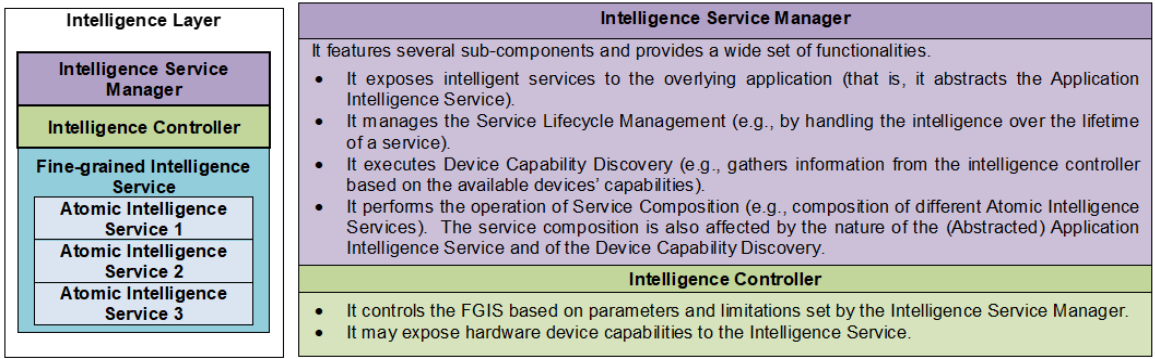

Fig. 6: Intelligence Layer key components high-level description

*D. Binding Applications and Intelligent Services*

As explained before, the intelligence layer delivers an intelligent service to any application that has requested it. In practical terms, the intelligent algorithm's output must comply with context and semantics that the application can interpret and use in its own functions. This compliance can be ensured by allowing the processed intelligence to access to the application through a microservice-based approach [20]. Specifically, the application layer gets access to a dedicated API to connect to the output delivered by the intelligence layer. When defining such an API, additional aspects must be considered. For example, many use cases include the execution of several intelligent applications. Additionally, such use cases rely on systems composed of several subsystems, which in turn respond to specific functionality and requirements for appropriate processing, input, and actions. A clear example of system complexity is given by a modern car. In a car, there are platforms dedicated for managing engine and mechanical systems, while others focus on the passenger's entertainment and comfort. Although such subsystems belong to the same car's integral system, each must interpret very different input and output information and processing requirements.

Regardless of this specific example, the way in which intelligence is handled can differ from case to case. A single platform could serve multiple applications where several intelligence providers may upload intelligent services. Or intelligent services could be redundantly distributed across multiple platforms, such as in cloud or local execution environments, to optimize their execution. Furthermore, another factor that must be considered is whether a single intelligent service can be executed independently or if it presents specific dependencies from other services.

## IV. Intelligence Provisioning and Orchestration

*A. Machine Intelligence Suppliers*

*1) Intelligence Service Providers:* Generic intelligence, or an intelligence that can perform any intellectual task at hand, is very difficult to achieve. The best intelligence for a specific task is perceived to come from experts and their specialized experience on the area of the task. This rule applies equally for MI and humans. Therefore, as the applications fields of MI widen, also the required specialization for the data and algorithms increases. From a techno-economic point of view, it is more optimal for intelligence service providers to specialize in a sub-area of expertise. This specialization means that intelligence service providers are focused on two areas: harvesting data and creating intelligent models. These intelligent models can later be distributed to local processing places or kept it in a centralized processing area. The intelligence service provider's role is directly about handling intelligence and it does not need to take care of the application making use of the intelligence. Therefore, intelligence service providers can primarily direct their resources into producing intelligence. The intelligence's know-how and data become an asset that requires a terms-of-use agreement with the end-user and confidentiality protections that must be enforced through security protocols and special suitable encryption solutions. The training data is also a valuable asset, since the collection of relevant data has been always proven difficult and the availability restricted. In this respect, it is reasonable to assume that certain types of intelligence providers become data providers. They may even only provision specialized annotated data that can be input to other models (for training or knowledge base purposes) or to support inference.

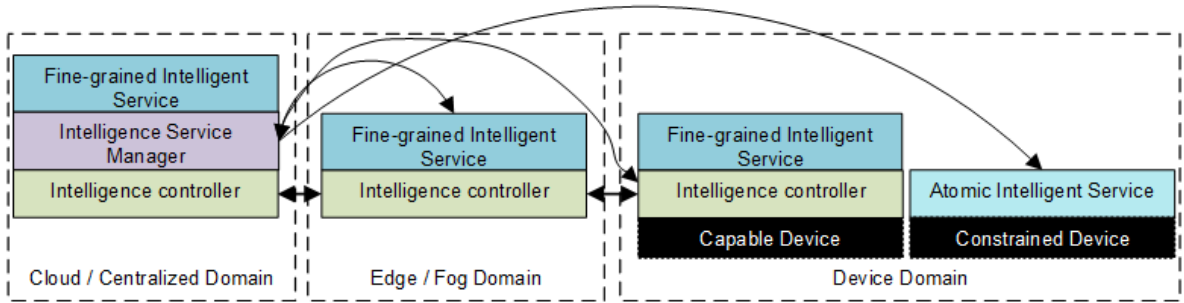

Fig. 7: Example of Interactions between Intelligent Layers from several domains in a specific IoT deployment

*2) Industrial Intelligence Service Providers:* Intelligence setup can be provided by the system itself or by an external entity to the system, similarly to the previously introduced Intelligence Service Provider. However, in different contexts such as the industrial domain, there are characteristics that make them differ respect a consumer domain. The main difference arises from the higher coupling of software and hardware in the industrial systems and the liabilities proper of commercial industrial agreements. These differences result in one special category of providers for this environment: the Industrial Intelligence Service Providers.

In some cases, the Intelligence Service Providers are internal entities that oversee intelligence distribution or update the edges that have been centrally collected by the organization's agents. Industrial intelligence service providers may supply one or several intelligent components of a more complex system. Occasionally, they do not have fully direct access to such components. In these instances, mediator systems may be present to act as gateways, proxies, or even as an intelligence distribution node to the target intelligence nodes and their distributed intelligent functions. An intelligence distribution node would then act as a facade that provides industrial intelligence service providers with a view of a complete system abstraction, thus enabling intelligence layer configuration.

*B. Devices, Nodes and Platforms*

According to the definition of intelligent devices provided in Figure 4, intelligence providers can place intelligent services regardless of the underlying hardware platform. Consequently, in order to be able to uniformly handle the pointed out diversity, there is need for some standardization on how the MI is provisioned to the intelligence devices.

From a device perspective, the practical advantage of decoupling intelligence from the application becomes clearer whenever the intelligence provider decides to enhance or change the device's intelligence capabilities, for example by coupling a new intelligence feature with one that is already hard-coded into the device. The intelligence provider has the flexibility to generate different fine-grained intelligence services by composing atomic-intelligence services. The main advantage of such capability lies in the fact that this kind of deployment can facilitate intelligence interoperability among devices and partially mitigate strict dependency on the hardware platform. In order to be compliant with the aforementioned specification, the platform implementation must ensure that the interface setup allows intelligence processing execution optimization, which in practice means mapping it to the most suitable code for a given platform. The last feature becomes extremely useful if we consider the high heterogeneity of hardware environments in which the intelligence must be deployed.

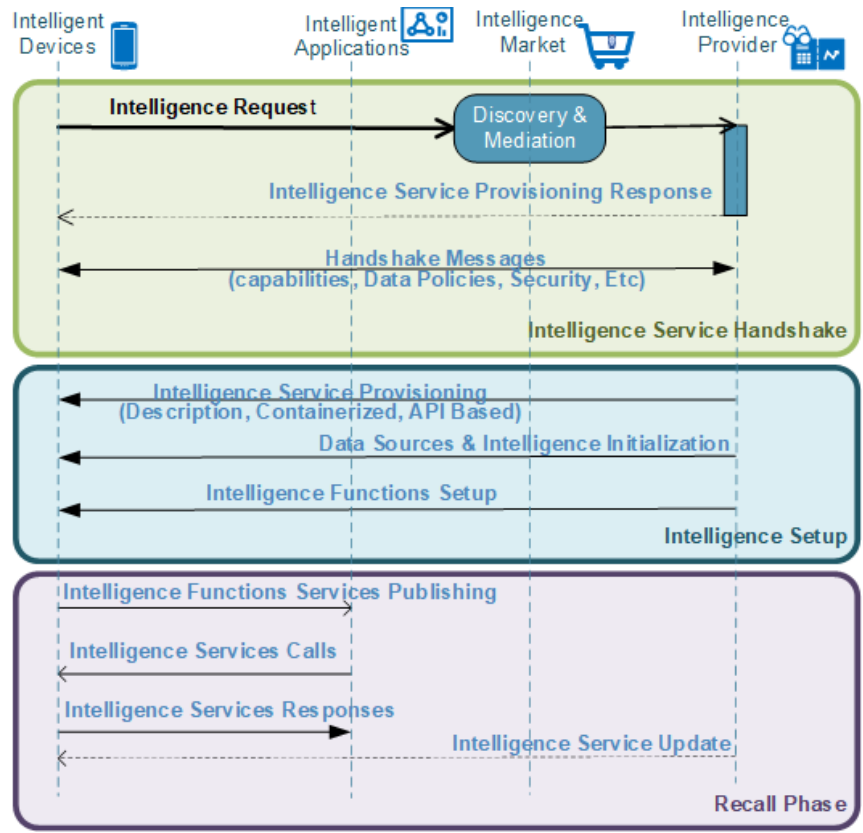

Fig. 8: Intelligence distribution framework in a consumer environment

On the basis of the above considerations, the mentioned interface must be composed by open or standardized API, as well as specific protocols that are able to interface any type of application and interact with any intelligence provider.

In the industrial context, the overall complexity of the system architecture increases. The higher presence of legacy components necessitates an additional ecosystem element in the device space called intelligent node. In contrast to intelligence devices, intelligent nodes can be defined as a higher-level abstraction of the components that belong to an already operative MI structure. There are three different base scenarios for intelligence nodes systems. The first scenario encompasses several components that highly depend on each other and that need a centralized control to cooperate. The second scenario embodies agents that are able to cooperate with peer components without centralized orchestration. Finally, the third scenario features multiple stand-alone agents that execute intelligence modeling collaboration without coordination (e.g. swarm intelligence), but it might also turn into competition. It is important to note that additional combinations of such system's models are possible, especially when they need to meet specific case studies' specifications. In today's modern industrial systems, the intelligence information exchange is left to application's implementation, which applies to all the previously mentioned models. For example, when the leanings of one node is propagated to all the others in a cooperative model, or when a central entity issues intelligence functions to the edges of the system, it is performed by applications own mechanisms. Also, in the stand-alone case, the intelligent function's update and configuration at on-boarding time, or during its operative lifetime, are also handled by application's implementations. Currently, in the industrial world, there is no clear dominant framework or standard that enables intelligence distribution; however, there have been attempts in this direction. For example, the Agent Communication Language (ACL) from the Foundation for Intelligent Physical Agents (FIPA) model was defined as a protocol for communication in multi-agent systems [21].

## C. Intelligence Provisioning Handshake

Intelligence's own nature and sensitivity requires more protection and handling care than, for example, displaying a web page. For intelligence, confidentiality and accessibility are the main concerns, therefore a more generic concept of privacy must be accounted for as well. Data sources are very valuable by themselves, not only because of the cost of acquiring high quality training data, but also for their significance in a larger context and in conjunction with correlations. For example, the information related to one car accident may not be of great concern for a car manufacturer, but a collection of data referring to all accidents involving one of their models might be. Machine learning algorithms' task is often finding correlations or models that can be used to make intelligent decisions. The resulting inferences may be of the same sensitive nature or even more than the data sources used by the intelligent functions. Today, depending on the existing regulatory framework, information provided or generated by applications could be exchanged between service providers and applications with a large diversity range of detail.

This information flow can be secured through terms-of-use agreements that give application users little or no information how the collected data is being used. Also, these agreements often do not provide ways for the users to restrict or discriminate when, what, and with whom information can be shared. The user's alternative could be just to quit or abandon the services, but in the case of very market dominant applications, it becomes difficult for users to find suitable fits for their needs. For data sets of high sensitivities such as customer records, intelligence service users require full data privacy coverage even if the data is processed outside of their premises or devices. This requirement implies that intelligence providers should be able to install machine learning algorithms without needing to know the actual values utilized as input. This process can be achieved with measures such as homomorphic encryption [22] or secure multiparty computations [23].

The intelligence providers may be also concerned about the confidentiality of the algorithms used from their intelligence harvesting, both from the perspectives of learning processes setup and recall phase parameters. Providers often wish to protect their internal knowledge as well as their research and development investments from intelligence copy attempts or unauthorized cloning. This issue becomes even more difficult to address when the attacks use MI to clone the knowledge [24] [25] and to put in place relevant protections is another area requiring more research.

*1) Intelligence Request and Capabilities Exchange:* Configuring and managing a D-MI requires high flexibility to address the existence of multiple types of devices, scenarios, services, and topologies of connectivity. Such flexibility is necessary from the very starting point when intelligence is first being requested. An initial exchange of capabilities and relevant characteristics between the target device and the intelligence providers allows providers to assess what type of algorithms and intelligence setup are proper and relevant for the device according to the application layer's expected services. During this exchange, the provider can evaluate if such provisioning for the requested intelligence on the target device is feasible or supported. The request process involves negotiation of data sharing policies and may also include charging aspects of the intelligence and data provisioning in any direction. The request can be done directly to an Intelligence Provider or through a mediator entity such as an Intelligence Market implementation. The market may use part of the information contained in the request to match the intelligence needs requested to the capabilities and services offered by the Intelligence Providers available in the market. This process of requesting the intelligence, exchanging capabilities, sharing policies and charging policies comprises an Intelligence Handshake.

Figure 8 displays an example of how different ecosystem entities interact amongst each other in order to enable intelligence provisioning. The different procedures are grouped according to some of the distinct phases of the MI LCM as defined in Figure 2.

*2) Sharing Policies:* sharing policies include what data can be shared and how, which services can access which data, and what data requires additional anonymization. Instead of being solely left to the parties involved, data sharing polices should be enforceable by protocols and processes introduced to set up the intelligence services. Otherwise, there is no warranty that a change of terms of service agreements would retroactively affect data exchanged under different agreed conditions. Profiles can be provisioned in multiple ways depending on the nature of the environment where the intelligence is executed. Ledgering technologies and smart contracts are two emerging tools that provide the means to audit and enforce data sharing policies that parties have agreed-upon [26] [27]. Such kind of tools equips the systems with capabilities to restrict data redistribution, grant several levels of data accessibility based on the agreed upon levels, and introduces the possibility of access revocation. These policies govern the relationships, permissions, and treatment of the data that is transmitted to and from intelligent providers.

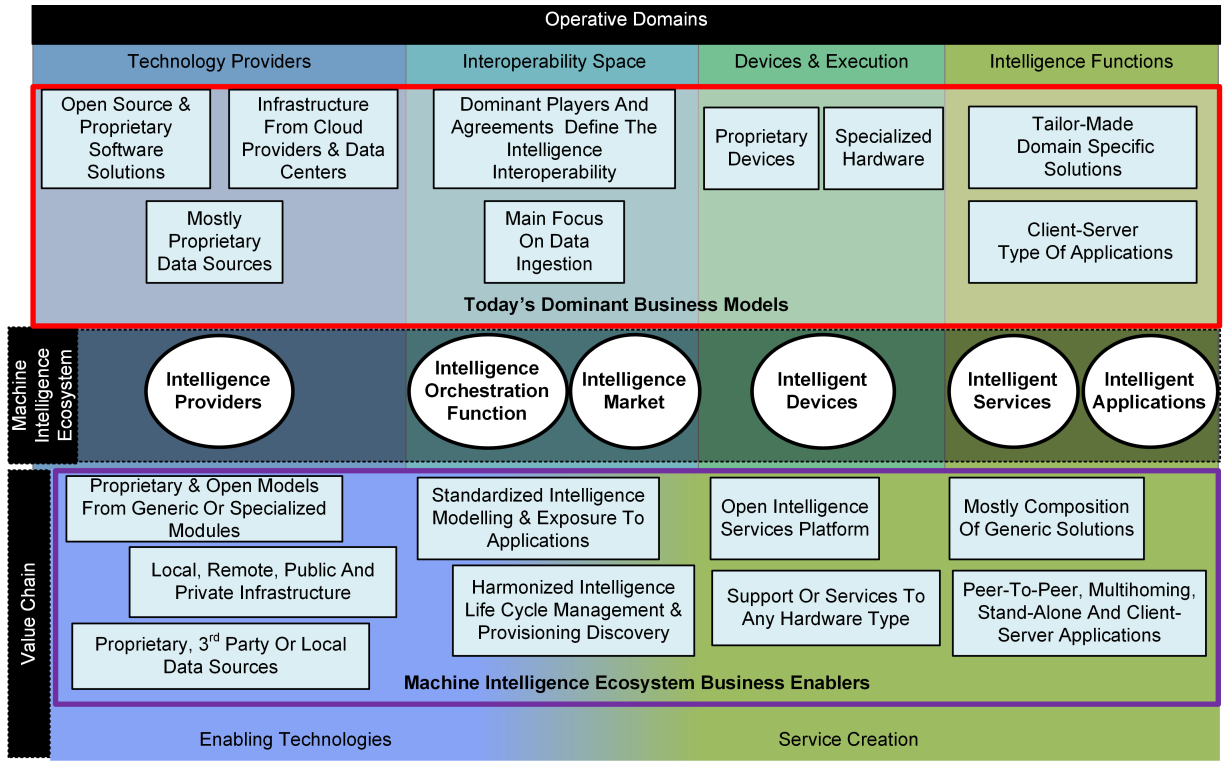

Fig. 9: Machine Intelligence Distribution value generation enablers

*D. Additional Technical Enablers and Challenges*

Technology is one the powerful forces for creating value in business ecosystems. Figure 9 illustrates the different domains that the MI ecosystem needs to address in order to provide intelligence in the IoT context. The technology drives the service creation that generates additional value to the current dominant business models. Therefore, several technological aspects must be in place in order to solve the challenges imposed by emerging new services.

An integral approach to security can not be ignored and it is important to protect intelligence users and intelligence service providers from fraud, privacy attacks, unauthorized accesses, confidentiality breaches, and sensitive personal and industrial data security. Establishing trust is an especially relevant challenge because of the framework's distributed nature; in other words, why and how to trust peers that are not known previously. Other challenges in addition to the aforementioned confidentiality and data access privileges are the misuse of intelligence for not agreed purposes, a lack of auditing tools, and traceability of MI decisions.

*1) The Intelligence Plane:* Intelligence providers must be able to define, configure, and initialize the intelligence functions from the intelligence layer. They can use several alternative approaches to either describe the intelligence or just implant it in the supporting platforms. A descriptive approach may use an intelligence descriptive language such as PMML[4] from the Data Mining Group or ONNX[5], which was created by Facebook and Microsoft. A descriptive language enables the mapping of abstraction entities and instructions of the language to algorithms and processes implemented in a platform supporting it. A descriptive language can be compared to a web browser that interprets a web page description file, displays the page according to its implementation in a specific platform, and makes use of any available acceleration hardware.

---

[4] http://dmg.org/pmml/v4-3/GeneralStructure.html/   [5] https://onnx.ai/

Another provisioning approach is to support and supply APIs by device platforms. Inter-vendor portability would need to cover a large diversity of device types and therefore, a number of capabilities. Despite the fact that this approach was proven successful with the adoption of GPUs for general computing, it may not be equally feasible when applied to devices with different machine learning capabilities translated in diverse types of processing hardware. Consequently, it would be difficult to define one architectural framework across devices, in contrast with the case of the GPUs which components and target function were less fragmented. APIs also tend to be proprietary, and are often extended to take full support of the available capabilities of the hardware or particular architecture, which leaves portability support to software programmers. TensorFlow [28] is an example of an open-source library that makes use platform API targeting to accelerate machine learning algorithms' numerical computation.

Another alternative is using pre-configured virtualized instances – ex. Virtual Machine (VM) or containers – that provide APIs to the application layer. The application layer can use this API to access to the intelligence from the algorithms and processes contained in the execution environment of the VM/container. In this last option, the intelligence layer implementation resembles a series of containers that each has its own API that is accessible by the application layer. This system would replace a more monolithic entity that provides a general intelligence service to higher layers. The pre-configured VM/container option would also present possibilities to procure independent and protected memory areas and restricted accessibility to the configuration and control interfaces.

*2) Semantic Interoperability:* An additional aspect to consider is the meaning of the data used for intelligence. The actual interpretation of a data set may vary between systems, which can lead to confusing or incorrect inferences. Semantic interoperability between the data models used by the data sources and the understanding of the data by the intelligence models is necessary and important. Current efforts in the IoT space, such as Light Weight Machine to Machine[6] (LWM2M) or World Wide Web Consortium (W3C) Web of Things[7], are working to provide frameworks, resources, and tools that potentially address not only the problem of understanding the nature of the data, but also to discover devices' capabilities. These initiatives describe possible interactions with the devices and how to understand their operational requirements and possible outputs. For intelligent devices and platforms, this knowledge also impacts the capabilities descriptions. A device should be able to unambiguously inform what intelligent actions and services it can provide, and its capabilities to support intelligent algorithms [29]. Additionally, semantic is needed to describe intelligent services to applications, and what are the required input and expected output that enables intelligence discovery. The goal is to provide a semantic description that should be understandable and operational for automatic intelligence configuration without human intervention.

## V. Conclusion

The spread of intelligence in all types of devices, from the consumer to industrial domains, requires specialized solutions that nowadays are provided by proprietary means (with very little interoperation) and supported by centralized systems. This tendency does not seem sustainable, especially when the number of devices and their capabilities are dramatically increasing. On the grounds of the presented limitations, we propose introducing an intelligence layer to the devices stack that allows multiple applications to benefit from the intelligence setup provided by intelligence providers. The introduction of an intelligence layer promotes an intelligence distribution ecosystem with an MI distribution framework operating on top of it. This ecosystem would enable the development of intelligence functions and processing models that are suitable to many MI use cases, including distributed, decentralized, or centralized instances. The framework outlines the relationship between entities and the provisioning of intelligence to the network edge in the IoT context. To realize the whole vision, it is important an effective integration of a set of technology enablers related to the control of the data flow, enforcing of the negotiated policies, and the possibility to change such agreements dynamically. These technology enablers and policies allow data users and providers to control and take full ownership of their data in any time. People must not be dependent only on one-time or unilaterally set agreements; they must have the freedom to access any intelligence services available.

---

[6] http://www.openmobilealliance.org    [7] https://www.w3.org/TR/wot-architecture/

When accounting for multiple deployment and applications scenarios, further definition of the intelligence layer components is needed. Moreover, harmonization of the MI LCM and discovery of the key integration points between applications and intelligence are research challenges that need to be further investigated. Moreover, the setup of data processing, from the source to their consumption point, requires additional considerations that might impact platform deployment and intelligent services execution. Connecting the intelligence layer processing with the relevant data sources is therefore one of the challenges of the described model.

# References

[1] D. Monroe, "Chips for artificial intelligence," *Commun. ACM*, vol. 61, no. 4, pp. 15–17, Mar. 2018.
[2] S. J. Russell and P. Norvig, *Artificial Intelligence: A Modern Approach*. Malaysia; Pearson Education Limited,, 2016.
[3] N. Heuveldop *et al.*, "Ericsson mobility report," *Ericsson AB, Technol. Emerg. Business, Stockholm, Sweden, Tech. Rep. EAB-17*, vol. 5964, 2017. [Online]. Available: https://www.ericsson.com/assets/local/mobility-report/documents/2017/ericsson-mobility-report.-november-2017.pdf
[4] C. MacGillivray, M. Torchia, M. Kalal, M. Kumar, R. Membrilla, A. Siviero, Y. Torisu, N. Wallis, and S. Chaturvedi, "Worldwide internet of things forecast update, 2016-2020," *IDC Research*, 2016.
[5] J. Gantz and D. Reinsel, "The digital universe in 2020: Big data, bigger digital shadows, and biggest growth in the far east," *IDC iView: IDC Analyze the Future*, vol. 2007, no. 2012, pp. 1–16, 2012.
[6] C. V. Networking, "Cisco global cloud index: Forecast and methodology, 2015-2020. white paper," *Cisco Public, San Jose*, 2016.
[7] C. Szegedy, W. Liu, Y. Jia, P. Sermanet, S. Reed, D. Anguelov, D. Erhan, V. Vanhoucke, and A. Rabinovich, "Going deeper with convolutions," in *2015 IEEE Conference on Computer Vision and Pattern Recognition (CVPR)*, June 2015, pp. 1–9.
[8] R. Girshick, I. Radosavovic, G. Gkioxari, P. Dollár, and K. He, "Detectron," https://github.com/facebookresearch/detectron, 2018.
[9] A. Torralba and A. A. Efros, "Unbiased look at dataset bias," in *CVPR 2011*, June 2011, pp. 1521–1528.
[10] B. McMahan, E. Moore, D. Ramage, S. Hampson, and B. A. y Arcas, "Communication-efficient learning of deep networks from decentralized data," in *Proceedings of the 20th International Conference on Artificial Intelligence and Statistics*, ser. Proceedings of Machine Learning Research, A. Singh and J. Zhu, Eds., vol. 54. Fort Lauderdale, FL, USA: PMLR, 20–22 Apr 2017, pp. 1273–1282.
[11] I. Stoica, D. Song, R. A. Popa, D. A. Patterson, M. W. Mahoney, R. H. Katz, A. D. Joseph, M. Jordan, J. M. Hellerstein, J. Gonzalez, K. Goldberg, A. Ghodsi, D. E. Culler, and P. Abbeel, "A berkeley view of systems challenges for AI," EECS Department, University of California, Berkeley, Tech. Rep. UCB/EECS-2017-159, Oct 2017.
[12] "Decentralized machine learning," White Paper, decentralizedml.com, Apr. 2018. [Online]. Available: https://decentralizedml.com/DML_whitepaper_31Dec_17.pdf
[13] B. Tang, Z. Chen, G. Hefferman, S. Pei, T. Wei, H. He, and Q. Yang, "Incorporating intelligence in fog computing for big data analysis in smart cities," *IEEE Transactions on Industrial Informatics*, vol. 13, no. 5, pp. 2140–2150, Oct 2017.
[14] H. Li, K. Ota, and M. Dong, "Learning IoT in edge: Deep learning for the internet of things with edge computing," *IEEE Network*, vol. 32, no. 1, pp. 96–101, Jan 2018.
[15] X. Chen, Q. Shi, L. Yang, and J. Xu, "Thriftyedge: Resource-efficient edge computing for intelligent iot applications," *IEEE Network*, vol. 32, no. 1, pp. 61–65, Jan 2018.
[16] Z. Chen and B. Liu, "Lifelong machine learning," *Synthesis Lectures on Artificial Intelligence and Machine Learning*, vol. 10, no. 3, pp. 1–145, 2016.
[17] D. L. Silver, Q. Yang, and L. Li, "Lifelong machine learning systems: Beyond learning algorithms," in *AAAI Spring Symposium: Lifelong Machine Learning*, vol. 13, 2013, p. 05.
[18] J. Zhao, T. Tiplea, R. Mortier, J. Crowcroft, and L. Wang, "Data analytics service composition and deployment on IoT devices," in *Proceedings of the 16th Annual International Conference on Mobile Systems, Applications, and Services*, ser. MobiSys '18. New York, NY, USA: ACM, 2018, pp. 502–504.
[19] O. Vermesan and P. Friess, *Internet of things: converging technologies for smart environments and integrated ecosystems*. River publishers, 2013.
[20] M. Fowler and J. Lewis, "Microservices, 2014," *URL: http://martinfowler. com/articles/microservices. html*, 2014.
[21] S. Poslad and P. Charlton, "Standardizing agent interoperability: The FIPA approach," in *ECCAI Advanced Course on Artificial Intelligence*. Springer, 2001, pp. 98–117.
[22] T. Graepel, K. Lauter, and M. Naehrig, "Ml confidential: Machine learning on encrypted data," in *International Conference on Information Security and Cryptology*. Springer, 2012, pp. 1–21.
[23] A. C. Yao, "Protocols for secure computations," in *Proceedings of the 23rd Annual Symposium on Foundations of Computer Science*, ser. SFCS '82. Washington, DC, USA: IEEE Computer Society, 1982, pp. 160–164.
[24] N. Papernot, P. McDaniel, I. Goodfellow, S. Jha, Z. B. Celik, and A. Swami, "Practical black-box attacks against machine learning," in *Proceedings of the 2017 ACM on Asia Conference on Computer and Communications Security*. ACM, 2017, pp. 506–519.
[25] F. Tramèr, F. Zhang, A. Juels, M. K. Reiter, and T. Ristenpart, "Stealing machine learning models via prediction APIs," in *USENIX Security Symposium*, 2016, pp. 601–618.
[26] R. Hull, V. S. Batra, Y.-M. Chen, A. Deutsch, F. F. T. Heath III, and V. Vianu, "Towards a shared ledger business collaboration language based on data- aware processes," in *International Conference on Service-Oriented Computing*. Springer, 2016, pp. 18–36.
[27] A. Ouaddah, A. A. Elkalam, and A. A. Ouahman, "Fairaccess: A new blockchain-based access control framework for the internet of things," *Security and Communication Networks*, vol. 9, no. 18, pp. 5943–5964, 2017.
[28] M. Abadi, P. Barham, J. Chen, Z. Chen, A. Davis, J. Dean, M. Devin, S. Ghemawat, G. Irving, M. Isard *et al.*, "Tensorflow: A system for large-scale machine learning," in *OSDI*, vol. 16, 2016, pp. 265–283.
[29] M. I. Robles, E. Ramos, N. Beijar, and N. C. Narendra, "Calculating LWM2M resource semantic distance through senact ontology," in *Proceedings of the Seventh International Conference on the Internet of Things*, ser. IoT '17. New York, NY, USA: ACM, 2017, pp. 12:1–12:8.